\documentstyle[aps,epsf]{revtex}
\begin{document}

\title{Spatiotemporal Dynamics of Mesoscopic Chaotic Systems}
\author{Raymond Kapral and Xiao-Guang Wu\\
Chemical Physics Theory Group,\\ Department of
Chemistry, University of Toronto,\\ Toronto, Ontario, Canada M5S 1A1}
\date{\today}
\maketitle

\begin{abstract} 
An investigation of the mesoscopic dynamics of chemical systems whose mass action 
equation gives rise to a deterministic chaotic attractor is carried out. A reactive 
lattice-gas model for the three-variable autocatalator is used to provide a 
mesoscopic description of the dynamics. The global and local dynamics is studied 
as a function of system size and diffusion coefficient. When the diffusion length 
is comparable to the system size phase coherence is maintained but the amplitudes 
of the oscillations are uncorrelated due to interaction between fluctuations and the 
instability of the chaotic dynamics. If the diffusion length is small compared to 
the system size then phase turbulence serves to destroy the noisy global attractor.
\end{abstract}

\section{Introduction}
The dynamics one observes on macroscopic scales has its origin in the molecular 
collision events that occur on the microscopic level. Under far-from-equilibrium 
conditions even the macroscopic dynamics can take complex forms and show chaotic or 
turbulent behavior. Macroscopic chaotic or ordered structures arise because the 
correlations extend to macroscopic scales. 
A complete understanding or description of the dynamics entails 
an examination of the system on the microscopic,  or at least mesoscopic, length 
scales where internal fluctuations enter naturally in the dynamical description.

When the macroscopic deterministic dynamics is chaotic it is perhaps even more 
intriguing to consider how such structured chaotic motion for the macroscopic 
fields arises from the apparently random collision processes at 
smaller scales. We are then led to  
consider the interactions of fluctuations on small scales with the
intrinsic instability of the chaotic dynamics on the macroscopic scales.

In this article we use a reactive lattice-gas automaton to provide a mesoscopic 
description of the reaction dynamics. We investigate a specific mass action 
chemical scheme, the three-variable autocatalator, which shows a period-doubling 
cascade to a chaotic attractor. The reaction scheme and its bifurcation 
structure are described in Sec.~\ref{model}.  
The lattice-gas model naturally incorporates
local particle number fluctuations as a result of random local reactive events and 
particle motion. Section~\ref{lattice} provides details of the construction of 
the automaton dynamics and demonstrates that its mean field limit is the 
mass action rate law. The spatiotemporal dynamics in the chaotic regime is 
the subject of Sec.~\ref{spatio}. The roles of internal fluctuations, diffusion  
and system size in determining the global and local dynamics are explored in 
this section. Section~\ref{scales} presents a discussion of the automaton 
space and time scales and 
their relation to those in physical systems, while Sec.~\ref{conc} contains the 
conclusions of the study.

The automaton dynamics considered here not only provides a more fundamental  
description of far-from-equilibrium, chaotic, chemical systems but also constitutes 
a scheme for the study of such systems in the mesoscale domain where standard 
reaction-diffusion equation models may be inappropriate.

\section{Three-Variable Autocatalator Dynamics} \label{model}
In order to investigate the mesoscopic dynamics of systems exhibiting 
deterministic chaos it is necessary to consider reaction mechanisms with at 
least three species, apart from pool chemicals whose concentrations are maintained 
at fixed values by external feeds and constrain the system to lie far 
from equilibrium. It is also important that chemical concentrations remain 
positive and follow mass-action kinetics.\footnote{One of the first models of this type to be constructed is 
that due to Willamowski and R\"{o}ssler, Ref.~\protect\cite{wr}} 
In this article we consider the three-variable 
autocatalator\cite{auto} 
which has a rich phase space structure including periodic and 
chaotic attractors\footnote{This model is an extension of the 
two-variable autocatalator 
developed by Gray and Scott, Ref.~\protect\cite{gs}} 
and allows us to investigate 
certain aspects of the dynamics without unnecessary complications arising 
from bistability in the relevant parameter range.\footnote{In the 
Willamowski-R\"{o}ssler (WR) model the chaotic attractor coexists 
with a stable fixed point which can lead to interesting noise-induced transition 
processes but whose description can complicate the interpretaion of internal 
noise effects on chaotic dynamics. Studies of noise-induced transition 
phenomena in the WR model can be found in Ref.~\protect\cite{bis}}

\begin{figure}[htbp]
\begin{center}
\leavevmode
\epsfxsize=3.2true in
\epsffile{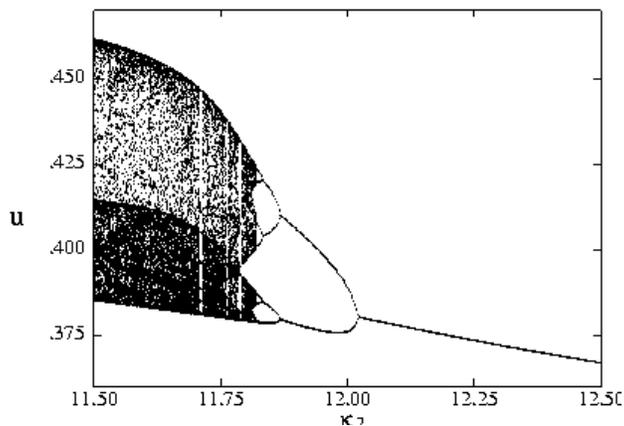}
\end{center}
\caption{Bifurcation diagram constructed from the solution 
of the mass-action equations (2). Here $u$ and $w$, respectively, 
are $U$ and $W$ concentrations on a Poincar\'{e} section 
$P$, defined as $\{P:\, \rho_v=v_{\rm s},\, \forall \rho_u,\rho_w \geq 0\}$. 
The bifurcation diagram shows the projection onto $u$ in the Poincar\'{e} 
section.}
\label{bidiag} 
\end{figure}

The three-variable autocatalator reaction mechanism is as
follows\cite{auto}:
\begin{equation}
\begin{array}{rcl}
A&\stackrel{k_0}{\longrightarrow}&U\\
A+W&\stackrel{k_1}{\longrightarrow}&U+W\\
U+2V&\stackrel{k_2}{\longrightarrow}&3V\\
U&\stackrel{k_3}{\longrightarrow}&V\\
V&\stackrel{k_4}{\longrightarrow}&W\\
W&\stackrel{k_5}{\longrightarrow}&B \;.
\end{array}\label{mechan}
\end{equation}
The chemical dynamics is described by the variations in the $U$, $V$ and $W$ 
concentrations, while concentrations of $A$ and $B$ are taken to be 
fixed by feeds of these species; consequently thier concentrations can be 
regarded as bifurcation parameters.
The mass action rate law corresponding to this mechanism is
\begin{eqnarray}
{d{\rho}_u \over dt}&=&\kappa_0+\kappa_1\rho_w-\kappa_2\rho_u\rho_v^2-\kappa_3\rho_u\, ,
\nonumber\\
{d{\rho}_v \over dt}&=&\kappa_2\rho_u\rho_v^2+\kappa_3\rho_u-\kappa_4\rho_v\, ,
\label{massact}\\
{d{\rho}_w \over dt}&=&\kappa_4\rho_v-\kappa_5\rho_w\, .\nonumber
\end{eqnarray}
Here $\rho_{\tau}(t)$ is the concentration of species 
$\tau=u,v,w$. The $\kappa_i$, $i=1, \cdots, 5$, are effective rate
constants that contain the concentrations of pool species. 
The system (\ref{massact}) has a steady state 
$(\rho_u,\rho_v,\rho_w)=(u_{\rm s},v_{\rm s},w_{\rm s})$
where $u_{\rm s}=\kappa_4v_{\rm s}/(\kappa_2v^2_{\rm s}+\kappa_3)$,
$v_{\rm s}=\kappa_0\kappa_5/\kappa_4(\kappa_5-\kappa_1)$ and 
$w_{\rm s}=\kappa_4v_{\rm s}/\kappa_5$.
This steady state may be destabilized by a Hopf bifurcation at which a
small-amplitude, oscillatory state emerges. In this study we choose
$\kappa_2$ as the bifurcation parameter while the values of the other 
parameters are 
fixed at $\kappa_0=1.666667$, $\kappa_1= 0.153$,  $\kappa_3=0.02$,
$\kappa_4=4.0$ and $\kappa_5=1.0$. With this choice of system parameters the
Hopf bifurcation point occurs at $\kappa_2=\kappa_2^{\rm H}=15.68$. The 
steady state is stable for $\kappa_2>\kappa_2^{\rm H}$. When $\kappa_2$ 
is decreased a period doubling cascade leading to chaos is observed 
(cf. Fig.~\ref{bidiag}). 

\section{Reactive Lattice-Gas Model} \label{lattice}
The mesoscale description of the autocatalator is based on a
reactive lattice-gas model of the dynamics. The
method of construction of such models along with specific applications
have been described earlier\cite{prepts,lgap} and, consequently, only sketch 
of those features of the model that pertain to the present application will
be given.  

Imagine that each chemical species 
$X_{\tau}$ ($U=X_1$, $V=X_2$, $W=X_3$) resides on its own lattice
${\cal L}_{\tau}$.  Each node of the species lattice ${\cal L}_{\tau}$ 
contains channels corresponding to unit velocities directed along the
links of the lattice from a node to its nearest 
neighbors.\footnote{One may also include zero velocity or stop particle 
channels 
in the node description (cf. Ref.~\protect \cite{prepts}).}
The nodes of the species lattices have identical node labels ${\bf r}$ 
so that several different species may reside at the same
spatial point in the system. There is no exclusion principle 
in the automaton model used in this study; $i.e.$, each velocity direction 
of a node on ${\cal L}_{\tau}$ may be occupied by any number of molecules 
$\alpha_{\tau}\geq 0$.\footnote{See, Refs.~\protect \cite{prepts} and 
\protect\cite{exclusion}. In the 
actual implementation of the model the occupancy of a node is restricted 
to some large number of particles, much larger than the average occupancy 
of a node. This insures that the elements of the reaction probability 
matrix always lie between zero and one.} 

Particle diffusion occurs through successive applications of
deterministic propagation of particles in directions determined by their 
velocities to neighboring lattice nodes, and velocity randomization where 
velocities are shuffled and particles are assigned new velocities at
random. We denote the combination of propagation and velocity
reandomization on ${\cal L}_{\tau}$ by $\hat{{\cal D}}_{\tau}$, the diffusion 
operator.

Consider a lattice ${\cal L}$ with ${\cal N}$ nodes labeled 
$({\bf r}_1,{\bf r}_2,\cdots,{\bf r}_{{\cal N}})$ and suppose many 
applications of the propagation and velocity randomization steps 
$\hat{{\cal D}}$ have been carried out. 
When there is no restriction on the number of particles at a node on a
lattice, the equilibrium distribution of $n$ particles on ${\cal N}$ 
nodes is an occupancy problem. The probability of obtaining occupancy numbers 
$\alpha({\bf r}_1),\alpha({\bf r}_2),\cdots,
\alpha({\bf r}_{\cal N})$ 
($\alpha({\bf r}_1)+\alpha({\bf r}_2)+\cdots+\alpha({\bf r}_{\cal N})=n$) 
on ${\cal N}$ nodes is given by \cite{kalbfleisch,feller}
\begin{equation}
P(\alpha({\bf r}_1),\alpha({\bf r}_2),\cdots,\alpha({\bf r}_{\cal N}))=
\frac{n!}{\alpha({\bf r}_1)!
\alpha({\bf r}_2)!\cdots \alpha({\bf r}_{\cal N})!}
{\cal N}^{-n}\, .\label{occup}
\end{equation}
The multimomial distribution (\ref{occup}) can be generated by
the multinomial theorem
\begin{equation}
\left(p_1+p_2+\cdots+p_{\cal N}\right)^{n}=
\sum_{\alpha({\bf r}_1),\alpha({\bf r}_2),\cdots,\alpha({\bf r}_{\cal N})}
P(\alpha({\bf r}_1),\alpha({\bf r}_2),
\cdots,\alpha({\bf r}_{\cal N}))\, ,
\end{equation}
with $p_1=p_2=\cdots=p_{\cal N}=1/{\cal N}$.
To obtain the particle number distribution on a single node we sum over all
the probabilities for the remaining ${\cal N}-1$ occupancy numbers:
\begin{equation}
P(\alpha({\bf r}_1))=\sum_{\alpha({\bf r}_2),\cdots,\alpha({\bf r}_{\cal
N})}P(\alpha({\bf r}_1),\alpha({\bf r}_2),\cdots,\alpha({\bf r}_{\cal N}))\, .
\end{equation}
The constraint on occupancy numbers $\alpha({\bf r}_2),\cdots,
\alpha({\bf r}_{\cal N})$ in the sum is 
$\alpha({\bf r}_2)+\cdots+\alpha({\bf r}_{\cal N})=n-\alpha({\bf r}_1)$ 
with $0\leq \alpha({\bf r}_1)\leq \alpha({\bf r}_{\cal N})$. Therefore we
obtain
\begin{equation}
P(\alpha({\bf r}_1))=\frac{n!}{(n-\alpha({\bf r}_1))!\alpha({\bf r}_1)!}
\left(\frac{1}{\cal N}\right)^{\alpha({\bf r}_1)}
\left(1-\frac{1}{\cal N}\right)^{n-\alpha({\bf r}_1)}\, ,
\end{equation}
where we have used the fact that 
\begin{equation}
\sum_{\alpha({\bf r}_2),\cdots,\alpha({\bf r}_{\cal N})}
\frac{(n-\alpha({\bf r}_1))!}
{\alpha({\bf r}_2)!\cdots \alpha({\bf r}_{\cal N})!}
{\cal N}^{-(n-\alpha({\bf r}_1))}=
[({\cal N}-1)/{\cal N}]^{n-\alpha({\bf r}_1)}.
\end{equation}

Since diffusion is carried out independently on all species lattices 
${\cal L}_{\tau}$, we may simply append the label $\tau$ to the 
node occupancy numbers and suppress the node label since all nodes 
are equivalent. Thus the particle number distribution at a node on 
${\cal L}_{\tau}$ is binomial:
\begin{equation}
P_{\rm b}(\alpha_{\tau})=\frac{n_{\tau}!}{(n_{\tau}-\alpha_{\tau})!\,
\alpha_{\tau}!}
\left(\frac{1}{\cal N}\right)^{\alpha_{\tau}}
\left(1-\frac{1}{\cal N}\right)^{n_{\tau}-\alpha_{\tau}} 
\, ,\label{bin}
\end{equation}
where $n_{\tau}$ is the total number of particles on ${\cal L}_{\tau}$. 
If we let ${\cal N} \rightarrow\infty$ and $n_{\tau}\rightarrow\infty$ 
in such a way that the ratio
$\rho_{\tau}=n_{\tau}/{\cal N}$ remains finite, the binomial distribution 
$P_{\rm b}(\alpha_{\tau})$ tends to the Poisson 
distribution\cite{kalbfleisch,feller}:
\begin{equation}
P_{\rm p}(\alpha_{\tau})=
\frac{(\bar{\rho}_{\tau})^{\alpha_{\tau}} }{\alpha_{\tau}!}
\exp(-\bar{\rho}_{\tau})\, .\label{poisson}
\end{equation}

The automaton reactive dynamics at a node is controlled by the reaction 
operator $\hat{{\cal C}}$ whose action is encoded in the 
probability matrix $P(\mbox{\boldmath$\alpha$}|\mbox{\boldmath$\beta$})$.
Each element of $P(\mbox{\boldmath$\alpha$}|\mbox{\boldmath$\beta$})$ gives
the transition probability from a reactant particle
configuration
$\mbox{\boldmath$\alpha$}=(\alpha_1,\alpha_2,\cdots,\alpha_s)$ 
to a product particle configuration
$\mbox{\boldmath$\beta$}=(\beta_1,\beta_2,\cdots,\beta_s)$, where 
$s$ is the number of species. If reaction is a rare
event and diffusion is able to maintain spatial homogeneity over the 
lattice the local particle distribution will be homogeneous 
(binomial or Poisson). In this case the (discrete) mean field rate equation 
for the concentration changes is easily written in terms of the reaction
probability matrix as
\begin{equation}
{\rho}_{\tau}(t+1)-{\rho}_{\tau}(t)=\sum_{\mbox{\boldmath$\alpha$}}
\sum_{\mbox{\boldmath$\beta$}}(\beta_{\tau}-\alpha_{\tau})
P(\mbox{\boldmath$\alpha$}|\mbox{\boldmath$\beta$}){P}_{\rm h}
\bigl(\mbox{\boldmath$\alpha$},\mbox{\boldmath$\rho$}(t)\bigr)\; ,
\label{meanfield}
\end{equation}
where $P_{\rm h}\bigl(\mbox{\boldmath$\alpha$},
\mbox{\boldmath$\rho$}(t)\bigr)=\Pi_{\tau}P_{\rm h}(\alpha_{\tau})
|_{\bar{\rho}=\rho_{\tau}(t)}$ with $P_{\rm h}(\alpha_{\tau})=
P_{\rm b}(\alpha_{\tau})$ or
$P_{\rm p}(\alpha_{\tau})$. In the limit $n_{\tau}\rightarrow\infty$ ($\forall
\tau$) and small reaction probabilities the mean field equation 
(\ref{meanfield}) should agree with 
the phenomenological mass action rate law (\ref{massact}). The reaction
probability matrix $P(\mbox{\boldmath$\alpha$}|\mbox{\boldmath$\beta$})$ 
can be constructed so that this mean field limit is obtained. If we restrict
particle number changes at a node to increases or decreases by one
particle on each
species lattice in accord with the mechanism of the reaction, one of the
possible choices for the reaction probability matrix is
\begin{equation}
\begin{array}{rcl}
P(\mbox{\boldmath$\alpha$}|\mbox{\boldmath$\beta$})&=&
p_1(\mbox{\boldmath$\alpha$})\delta_{\beta_u,\alpha_u+1}\delta_{\beta_v,
\alpha_v}\delta_{\beta_w,\alpha_w}
+p_2(\mbox{\boldmath$\alpha$})\delta_{\beta_u,\alpha_u-1}\delta_{\beta_v,
\alpha_v+1}\delta_{\beta_w,\alpha_w}\\
&+&p_3(\mbox{\boldmath$\alpha$})\delta_{\beta_u,\alpha_u}\delta_{\beta_v,
\alpha_v-1}\delta_{\beta_w,\alpha_w+1}
+p_4(\mbox{\boldmath$\alpha$})\delta_{\beta_u,\alpha_u}\delta_{\beta_v,
\alpha_v}\delta_{\beta_w,\alpha_w-1}\, ,
\end{array}\label{cmax} \end{equation}
where
\begin{eqnarray}
p_1(\mbox{\boldmath$\alpha$})&=&h(\kappa_0+\kappa_1\alpha_w)\, ,\nonumber\\
p_2(\mbox{\boldmath$\alpha$})&=&h[\kappa_2\alpha_u\alpha_v(\alpha_v-1)+
\kappa_3\alpha_u]\, ,\nonumber\\
p_3(\mbox{\boldmath$\alpha$})&=&h\kappa_4\alpha_v\, ,\nonumber\\
p_4(\mbox{\boldmath$\alpha$})&=&h\kappa_5\alpha_w\, .
\end{eqnarray}
Here $h$ is a factor that determines the time scale in the automaton
simulation. 
If (\ref{cmax}) is inserted into (\ref{meanfield}) the mass action rate law
(\ref{massact}) is recovered. Thus we may now investigate the full
automaton dynamics and compare its behavior to the autocatalator mass
action rate law and reaction-diffusion 
equation.\footnote{The reactive lattice-gas automaton bears some 
relation to birth-death master equation methods. 
See, for instance, Ref.~\protect\cite{master} }

In summary, the automaton dynamics is specified by the compositions of the 
diffusion $\hat{{\cal D}}_{\tau}$ and reaction $\hat{{\cal C}}$ opertors: 
\begin{equation}
\hat{T}=\prod_{\tau=1}^s\left(\hat{{\cal D}}_{\tau}\right)^{l_{\tau}}
\circ\hat{C}^{l_c} \, .\label{Tnr}
\end{equation}
In this equation $l_{\tau}$ and $l_c$ are non-negative integers, 
indicating that the operations are repeated $l_{\tau}$ and $l_c$ times, 
respectively. The reactive lattice gas model is a synchronously-updated,
probabilistic cellular automaton.

The simulations reported in this paper were carried out on triangular
lattices with ${\cal N}=N \times N$ nodes with periodic boundary 
conditions. On such triangular lattices the
application of the diffusion operator $\hat{{\cal D}}$ yields a
diffusion coefficient $D=1/4$ in units of lattice spacing squared per time
step. For the composition rule in (\ref{Tnr}) the diffusion coefficient 
of species $\tau$ is given by $D_{\tau}=(l_{\tau}/l_c) D$. In the 
simulations we have taken the diffusion coefficients to be equal: 
$D_u=D_v=D_w$. The time scale factor was taken to be $h=0.001$. 

\section{Spatiotemporal Dynamics in the Chaotic Regime} \label{spatio}
A complete understanding of the the dynamics of chaotic systems on
mesoscopic scales involves the consideration of a number of different
aspects. For systems which are sufficiently large and contain many
particles but are well-mixed, either through diffusion or mechanical
means, one expects that the chaotic attractor will resemble that of the
mass action rate law, apart from modifications of the 
attractor structure on small phase-space scales.\cite{wkref,wupre,nicolis}
If the system is large and molecular diffusion is the sole mechanism
responsible for eliminating concentration gradients in the system then the
full spatiotemporal dynamics must be considered. Even in such a
circumstance there is interesting information in the global
(spatially averaged) concentration fields and the attractors corresponding 
to these global variables; for example, for coupled map lattices composed 
of chaotic maps non-trivial, low-dimensional, global attractors have been
observed.\cite{chate} The nature of such global attractors has also 
been studied in lattices of diffusively coupled R\"{o}ssler oscillators
where the local R\"{o}ssler ODE dynamics is chaotic.\cite{bcm} 
The description of all aspects of the noisy
mesoscale dynamics involves a study of the spatial structure that
underlies the temporal evolution. This section is devoted to an
investigation of some of these issues for the autocatalator where 
we focus on the dynamics for $\kappa_2=11.52$, a parameter
value lying in the chaotic regime. 

\subsection{Noisy chaotic attractor: global dynamics}
We observed above that if the system is small enough so that diffusion 
is able to maintain spatial homogeneity over the entire lattice but 
contains a sufficient number of particles that fluctuations in the mean 
concentrations are small, the automaton attractor constructed from 
the globally-averaged concentrations will closely 
resemble that determined from a solution of the
mass-action ODE. Detailed studies of this type have been carried out for
the Willamowski-R\"{o}ssler model as a function of system size where the
automaton rule has been modified to guarantee perfect mixing.\cite{wupre} 
These calculations have confirmed that the gross structure of the chaotic 
attractor survives under the influence of internal noise, although noise
can lead to destruction of fine-scale structure (e.g., band merging in
banded attractors). There has been considerable discussion in the literature
about the effects of internal noise in chaotic systems and detailed
discussions can be found in Refs. \cite{wkref,wupre,nicolis,fox,chapter}.

\begin{figure}[htbp]
\begin{center}
\leavevmode
\epsfxsize=3.2true in
\epsffile{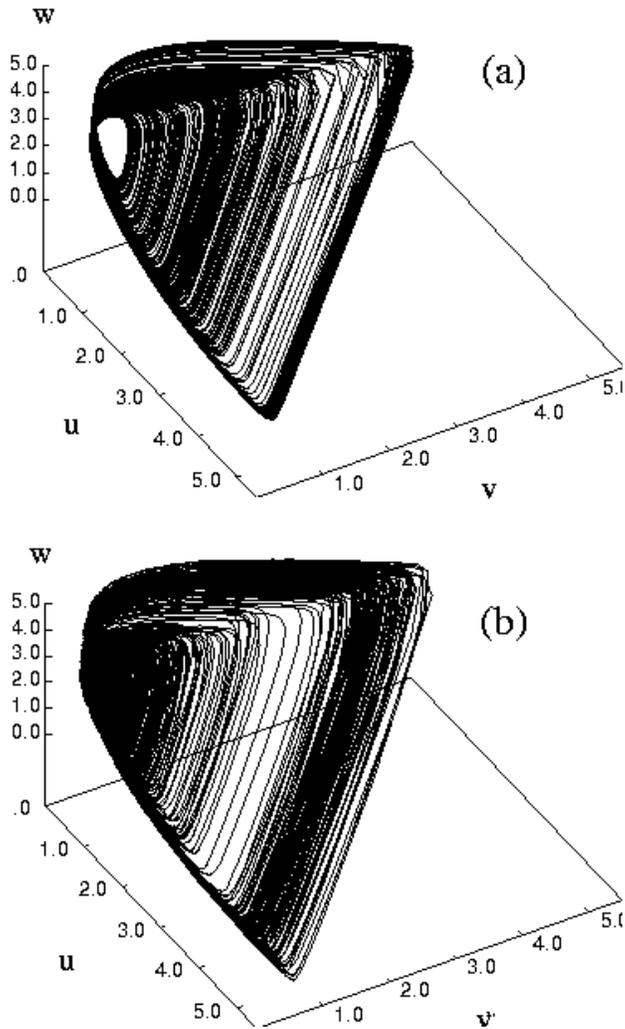}
\end{center}
\caption{ 3-D phase space plots of the chaotic attractor for
$\kappa_2=11.52$: (a) mass action rate law
(\protect\ref{massact}) and (b) automaton simulation on a 
$32 \times 32$ lattice.}
\label{attract}
\end{figure}

As an example, for a lattice 
of size $32 \times 32$ nodes we show that the noisy attractor 
(cf. Fig.~\ref{attract}(b)) 
constructed from the spatially-averaged concentrations,
\begin{equation}
\bar{\rho}_{\tau}(t)={N}^{-2}
\sum_{{\bf r} \in {\cal L}_{\tau}} \rho_{\tau}({\bf r},t)\;,
\end{equation}
resembles the deterministic attractor shown in panel (a) of
Fig.~\ref{attract}, although there are noticeable effects due to
fluctuations; as expected, the fluctuations tend to spread the dynamics in
the unstable directions and obliterate some of the attractor structure.
These effects diminsish as the system size increases provided spatial
homogeneity can be maintained.\cite{wupre}

\begin{figure}[htbp]
\begin{center}
\leavevmode
\epsfxsize=3.2true in
\epsffile{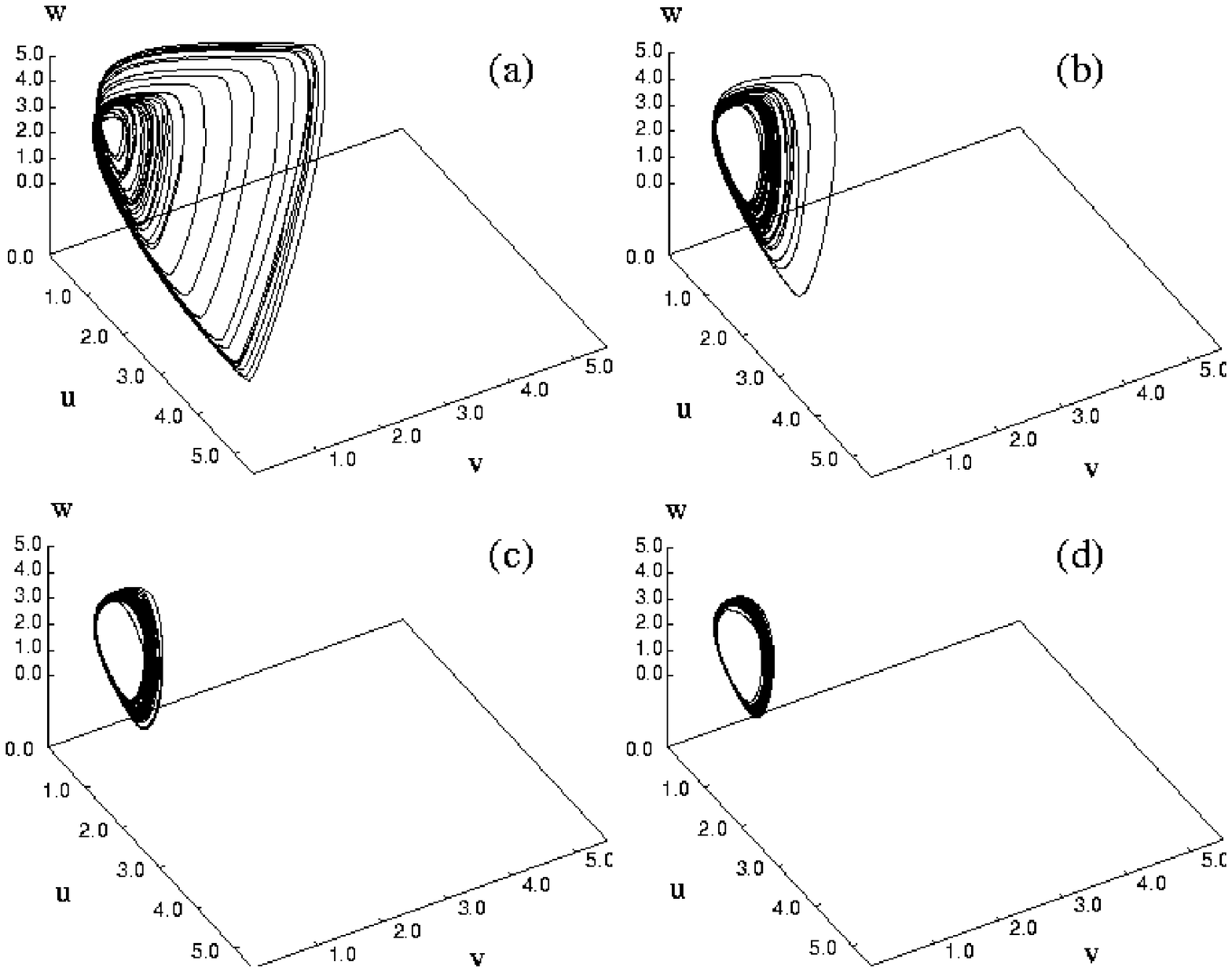}
\end{center}
\caption{ 3-D phase space plots of the global concentrations for
$\kappa_2=11.52$ for various system sizes: (a) $64 \times 64$ lattice, 
(b) $128 \times 128$ lattice, (c) $256 \times 256$ lattice and 
(d) $512 \times 512$ lattice.}
\label{sizes}
\end{figure}

Next, we consider the attractor structure determined from the
spatially-averaged concentrations as a function of the system size for a
fixed value of the diffusion coefficient, $D_{\tau}=1/4$. The  
results of a series of simulations on a set of $N \times N$ triangular
lattices, with $N=64,\;128,\;256$ and $512$, are shown in
Fig.~\ref{sizes}. One observes that as $N$ increases the global
attractor increasingly resembles a noisy periodic attractor whose width
shrinks as the system size increases, at least on the time scale of the
simulations. In the remainder of this section we examine the
structure of the spatiotemporal dynamics that underlies this behavior.

\subsection{Spatiotemporal structure}
\begin{figure}[htbp]
\begin{center}
\leavevmode
\epsfxsize=3.2true in
\epsffile{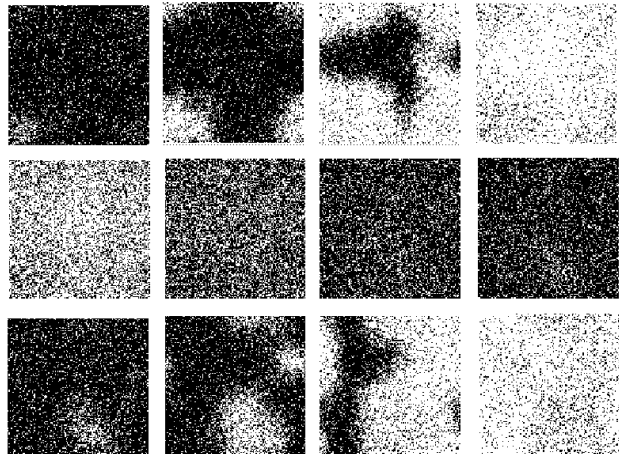}
\end{center}
\caption{The twelve panels of this figure show the local concentrations of
species $U$ for various times indicated in Fig.~\protect\ref{tseq}. Time 
increases from left to right and top to bottom and the total time spanned 
is about one and a half cycles on the attractor. The concentration is 
coded by gray shades with black high concentration and white 
low concentration of $U$.}
\label{sp128}
\end{figure}

\begin{figure}[htbp]
\begin{center}
\leavevmode
\epsfxsize=3.2true in
\epsffile{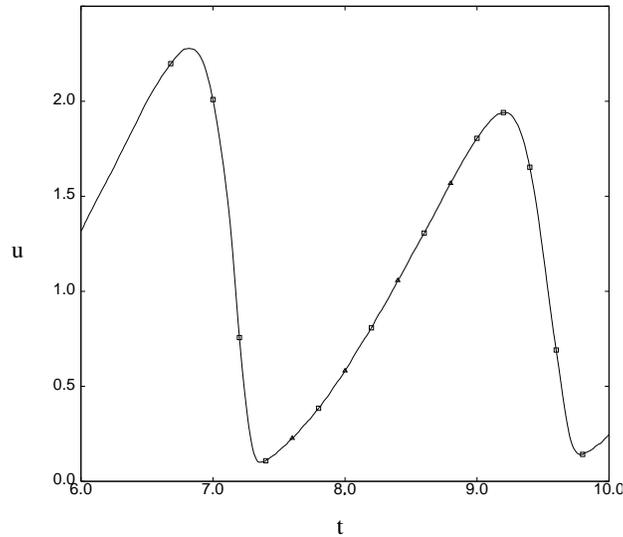}
\end{center}
\caption{Plot of the global concentration of species $U$ in the time
interval corresponding to the panels showing the spatial structure of the 
the $U$ concentration field. The points on the curve are taken at equally
spaced time intervals. The panels in Fig.~\protect\ref{sp128} correspond
to the open squares while the open triangles 
denote intermediate times where the structure is
not displayed.}
\label{tseq}
\end{figure}

We shall now examine in some detail the spatiotemporal dynamics on 
a $128 \times 128$
lattice. The spatial structure that underlies global attractor discussed
above is presented in Fig.~\ref{sp128}. This figure shows the
local concentration per node of species $U$ for different times 
(indicated by open squares) along a trajectory segment of the global 
attractor given in Fig.~\ref{tseq}.  
Strong concentration gradients develop during
the fast parts of the cycle while only small gradients exist during the
slow parts of the cycle. Of course, this behavior is anticipated in view
of the ability of diffusion to homogenize the concentration variations
that arise from local reaction. The slow part of the cycle, $t_s$, 
comprises most of the average cycle time on the attractor, $t_c$. For the
parameter value under consideration, $\kappa_2=11.52$, we have $t_c \sim
2.3$, while $t_s \sim 2.0$ and thus for the fast part of the cycle $t_f
\sim 0.3$. Consequently, the diffusion lengths corresponding to these
times are: $\ell_D^s \sim 32$ and $\ell_D^f \sim 12$ lattice sites, 
respectively. Since comparable concentration variations occur in the slow
and fast parts of the cycle diffusion will clearly be able to more
effectively homogenize the system in the slow part of the cycle than in
the fast part of the cycle. 

The spatiotemporal structure for the same parameters but on a 
$512 \times 512$ lattice is shown in Fig.~\ref{st512}. Once again the
strongest spatial inhomogeneities occur during the the fast part of the
cycle but even in the slow parts, as is evident from the first three
panels in the figure, one sees the small-amplitude inhomogeneities on 
the same length scale as in Fig.~\ref{sp128} more clearly since the
system size is so much larger than the diffusion length.

\begin{figure}[htbp]
\begin{center}
\leavevmode
\epsfxsize=3.2true in
\epsffile{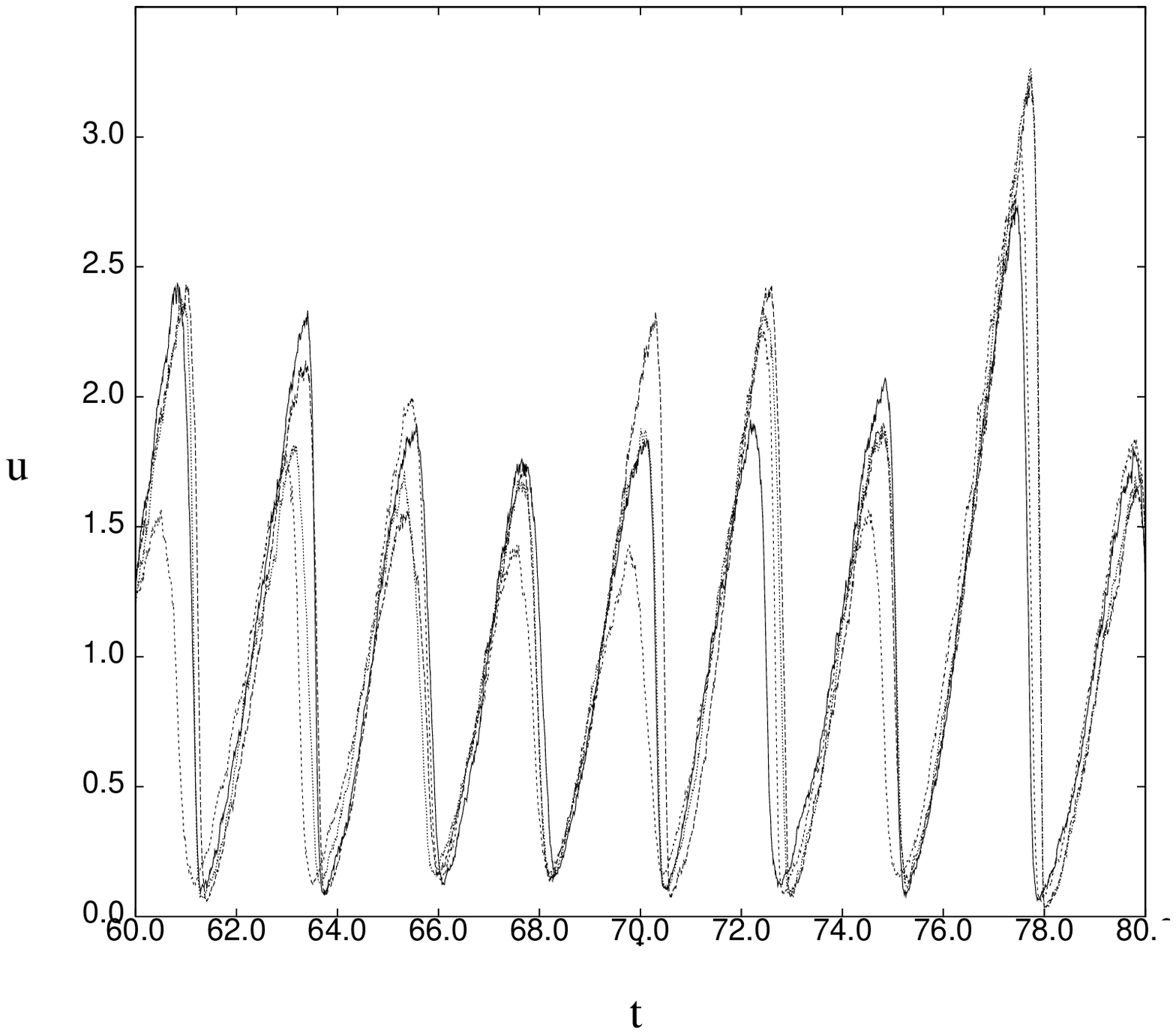}
\end{center}
\caption{Plot of the coarse-grained concentration of species $U$, 
$\bar{\rho}_{u}(\bar{{\bf r}},t)$,  in four cells as a function of time 
for $\kappa_2=11.52$.}
\label{coarse}
\end{figure}

The spatial structure may be examined in more detail by 
considering the coarse-grained dynamics of the system.
For this purpose we divide the system into cells with linear dimension $L$
and define the coarse-grained concentration fields as,
\begin{equation}
\bar{\rho}_{\tau}(\bar{{\bf r}},t)=L^{-2} \sum_{{\bf r} \in {\cal
L}_{\tau}(\bar{\bf r})} \rho_{\tau}({\bf r},t)\;,
\end{equation}
where ${\cal L}_{\tau}(\bar{\bf r})$ is a domain of the lattice centered
on $\bar{\bf r}$ containing $L \times L$ nodes.  The time variations of 
$\bar{\rho}_{u}(\bar{{\bf r}},t)$ for four cells with $L=32$ are shown in
Fig.~\ref{coarse}. The chaotic oscillations in the individual cells remain
in phase (at least over the time scale of the 
simulation\footnote{The total simulation time was 200 time units comprising
about 100 cycle times on the attractor.}) but their
amplitudes vary erratically, especially during the transition from the slow
part of the cycle to the fast part where fluctuations are strongly
amplified. Such regions of the
trajectory are expected to be most affected by noise arising from 
internal fluctuations.\footnote{Discussions of the effects of external noise on 
different parts of the cysle for both chaotic and periodic dynamics 
can be found in Ref.~\protect\cite{localf}}

Effects of this type were discussed earlier\cite{bohr} in a consideration 
of the global dynamics of coupled map lattices with chaotic or 
periodic local elements. In the chaotic regime
any internal fluctuation $\delta \rho_{\tau}({\bf r},0)$ will grow roughly as 
\begin{equation}
\delta \bar{\rho}_{\tau}({\bf r},n_{c}) = e^{\lambda t_{c} n_{c}} 
\delta \bar{\rho}_{\tau}({\bf r},0)\;,
\end{equation}
where $n_{c}=t/t_{c}$ is the number of cycles on the chaotic
attractor and $\lambda$ is the Lyapunov exponent. Here we have taken 
$\bar{\rho}_{\tau}(\bar{\bf r},t)$ to be the coarse-grained density.  
If the system is well-stirred the particle fluctuations at each node of
the lattice should be Poisson distributed as discussed in the previous
section. For finite diffusion, given the estimate of the diffusion length, 
one may expect Poisson-distributed fluctuations within the coarse-grained 
cells of size $L=32$. The Poisson fluctuations
within such coarse-grained regions provide a source of perturbations of
the dynamics of magnitude $\delta \bar{\rho}_{\tau}(\bar{\bf
r},0)=(\bar{\rho}_{\tau})^{1/2}/L$ which are then amplified by
the chaotic dynamics. One may then estimate the number of cycles on the
attractor for the system to lose memory of its initial state as 
\begin{equation}
n^*_{c} \sim {1 \over \lambda t_{c}} 
\ln {\delta \bar{\rho}^*_{\tau}(\bar{\bf r})L \over 
(\bar{\rho}_{\tau})^{1/2}}\;,
\end{equation}
where $\delta \bar{\rho}^*_{\tau}(\bar{\bf r})$ is a value of 
$\delta \bar{\rho}_{\tau}(\bar{\bf r},t)$ comparable to the ``size" of the
attractor in phase space. Using the cycle time on the attractor, $t_c
\equiv 2.3$,  and the fact that the Lyapunov exponemt is 
$\lambda =.229$ for $\kappa_2=11.52$ one finds that $n^*_{c}$
is of the order of a few cycles. Thus, given this time, one expects that
regions separated by distances of order $\ell_c \sim (2D t_{c}
n^*_{c})^{1/2}$ will have the amplitudes of their oscillations
uncorrelated. One finds $\ell_c  \sim 33$ and thus the amplitudes in the $32
\times 32$ coarse-grained cells should vary randomly as observed in the
simulations. If one then spatially averages over the coarse-grained cells
one should see noisy periodic dynamics since the random amplitudes will average 
to a nearly constant value. Note that phase coherence has not yet been 
destroyed and the above arguments apply to the structure transverse to 
the attractor. 

\begin{figure}[htbp]
\begin{center}
\leavevmode
\epsfxsize=3.2true in
\epsffile{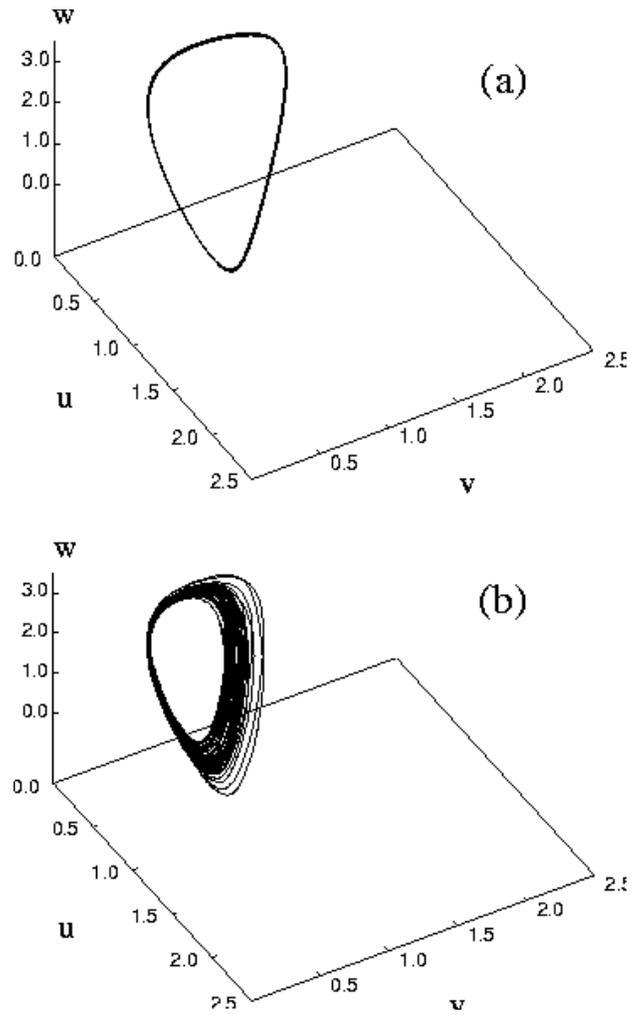}
\end{center}
\caption{Global phase space dynamics for $\kappa_2=12.8$, a parameter 
value in the periodic regime, for (a) the deterministic mass action 
rate law (\protect\ref{massact}) and (b) automaton dynamics on a 
$128 \times 128$ lattice.}
\label{osc128}
\end{figure}

\begin{figure}[htbp]
\begin{center}
\leavevmode
\epsfxsize=3.2true in
\epsffile{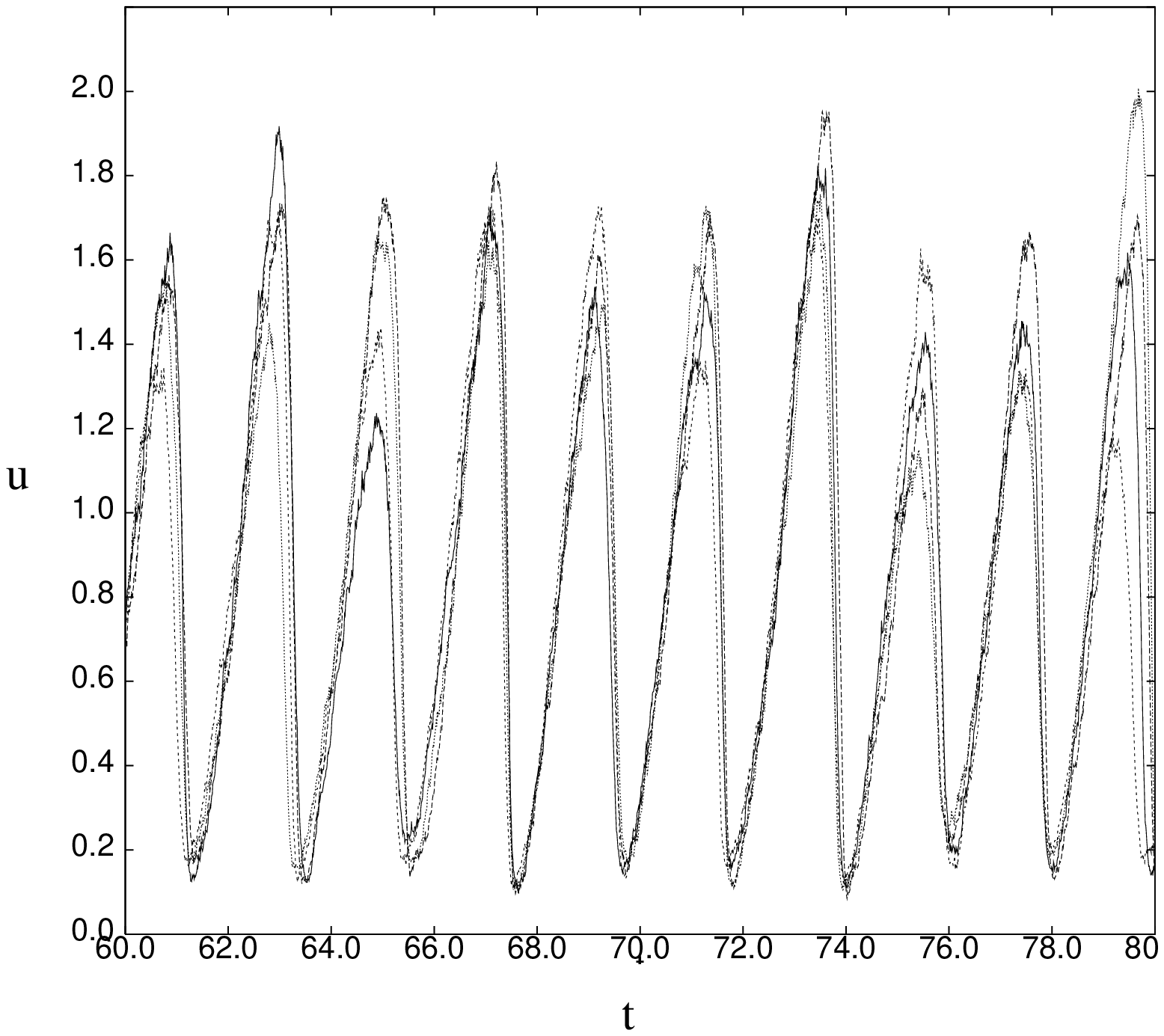}
\end{center}
\caption{Plot of the coarse-grained concentration of species $U$, 
$\bar{\rho}_{u}(\bar{{\bf r}},t)$, in four cells as a function of time 
for $\kappa_2=12.8$.}
\label{c128}
\end{figure}

It is interesting to compare the above dynamics with that in the periodic
regime. The results for $\kappa_2=12.8$ are shown in Fig.~\ref{osc128}
where panel (a) is the deterministic attractor and 
panel (b) is the attractor constructed from the global dynamics determined
from a  spatial average over the $128 \times 128$ lattice. As expected, 
a noisy periodic attractor is obtained with strong amplification of the
fluctuations in the fast part of the cycle. 
If one examines the coarse-grained
dynamics that underlies the noisy periodic orbit (cf. Fig.~\ref{c128}) one
sees local dynamics that is qualitatively similar to that for the chaotic
dynamics described earlier: the internal fluctuations most strongly 
influence the the portion of the cycle where the dynamics changes from 
slow to fast, and phase
coherence is maintained for this system size and value of the diffusion
coefficient. In the periodic regime, from a linearized analysis of the 
dynamics, one expects the fluctuations about the deterministic 
motion to grow linearly with time.\cite{fox,wkref} The fact that 
the dynamics observed in our simulations is similar in both the 
chaotic and periodic regimes likey reflects the small magnitude of the 
positive Lyapunov exponent for the autocatalator chaotic attractor. 
This implies that rather large spatial regions may remain correlated 
even in the chaotic regime. Similar persistence of correlations due 
to small positive values of the Lyapunov exponent were noted by 
Brunnet et al. \cite{bcm} in studies of coupled R\"{o}ssler ODEs.

In summary, the picture of the dynamics that is seen is consistent with
the qualitative bifurcation diagram in Bohr et al.\cite{bohr}; for the
finite-amplitude noise arising from internal fluctuations in the dynamics
one sees a noisy periodic orbit in both the chaotic and periodic regimes. 
For the relatively small lattice size, $128 \times 128$, the system 
maintains phase
coherence but noise interacts with the instability in the dynamics in 
directions along the unstable manifolds of the chaotic attractor to lead
to erratic amplitude oscillations whose spatial average gives rise to a
noisy periodic attractor.

\subsection{Destruction of phase coherence}
As the system size increases, or alternatively the diffusion coefficient
decreases, it will become impossible for the system to maintain phase
coherence over the entire spatial domain.\cite{kur,bennett} 
To investigate the passage to
this regime we again consider $128 \times 128$ lattices but vary the ratio 
$\ell_{\tau}/\ell_r$ to change the magnitude of the diffusion coefficient (and
reactive time scale). In the automaton simulations we increase the value
of $\ell_r$ keeping $\ell_{\tau}$ fixed so that the effective reaction rates
are increased. 

\begin{figure}[htbp]
\begin{center}
\leavevmode
\epsfxsize=3.2true in
\epsffile{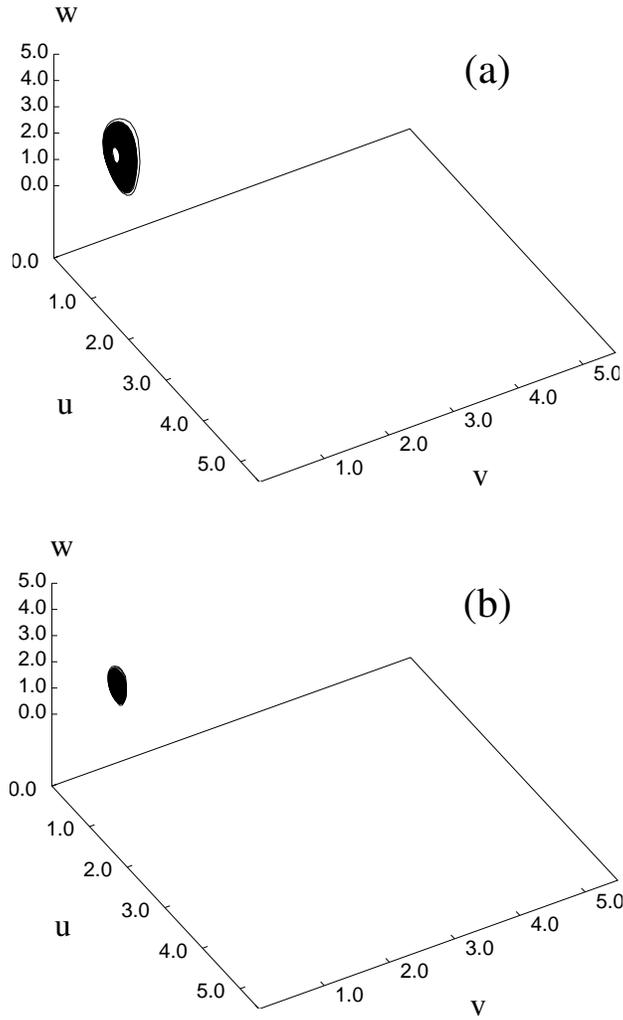}
\end{center}
\caption{Phase space plots of the global atractors determined from 
automaton dynamics on $128 \times 128$ lattices for two values of the
diffusion coefficient: (a) $D_{\tau}=D/6$ and (b) $D_{\tau}=D/10$.}
\label{bigd}
\end{figure}

\begin{figure}[htbp]
\begin{center}
\leavevmode
\epsfxsize=3.2true in
\epsffile{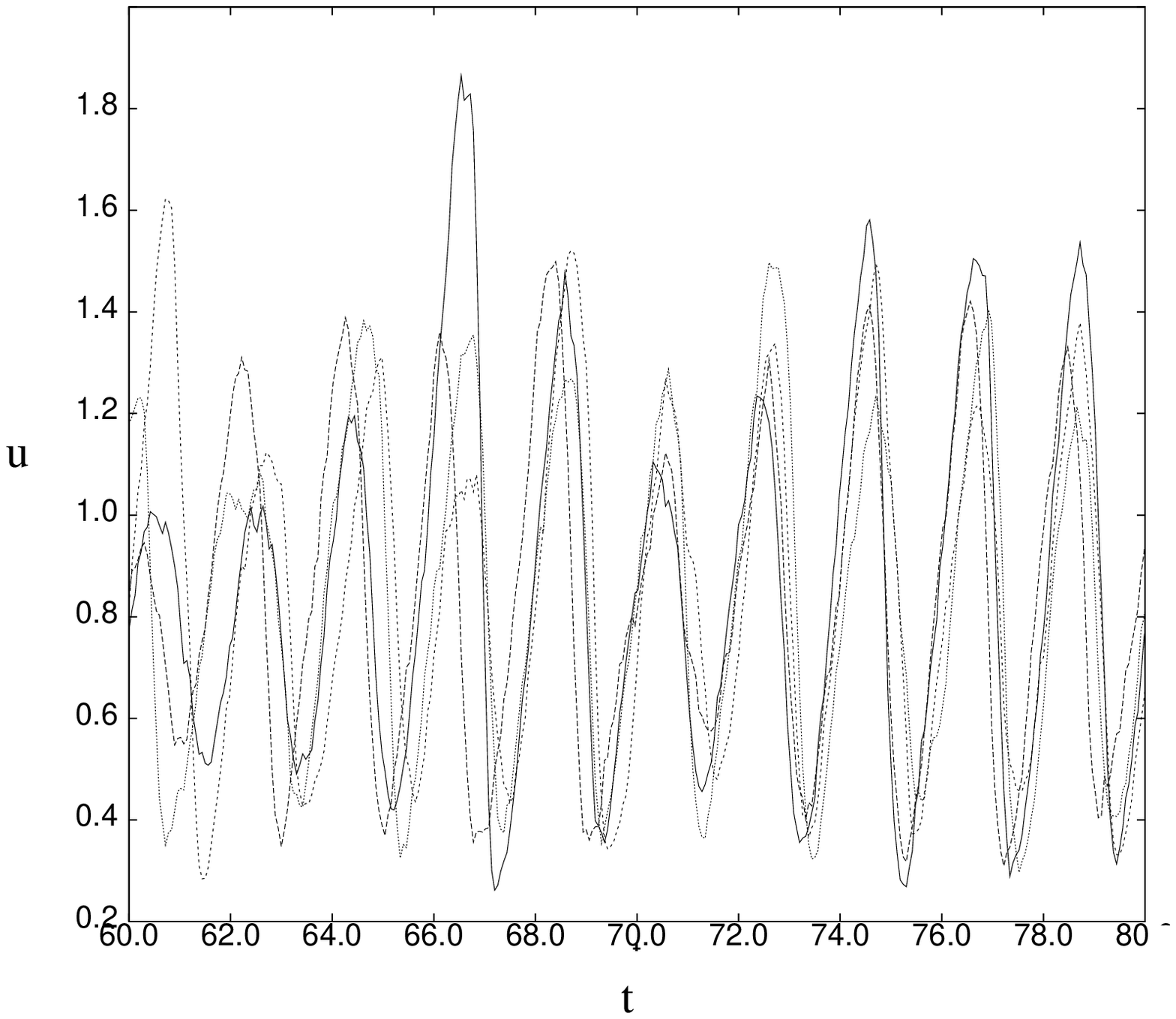}
\end{center}
\caption{Plot of the coarse-grained concentration of species $U$, 
$\bar{\rho}_{u}(\bar{{\bf r}},t)$, in four cells as a function of time 
for $\kappa_2=11.52$ for $D_{\tau}=D/6$.}
\label{corbd}
\end{figure}

The global attractors for $D_{\tau}=D/6$ and $D/10$ are shown in
Fig.~\ref{bigd}. They shrink to a noisy fixed point attractor as 
synchronization of the phase in space is lost. 
This is easily confirmed from an
examination of the coarse-grained dynamics in Fig.~\ref{corbd}: now ones
sees that both the amplitude and phase vary in the coarse-grained cells.

If the description of the dynamics were based on the reaction-diffusion
equation it would be possible to account for changes in the overall 
reaction rates and diffusion coefficients by scaling space and time. 
If such scaling were valid, one could relate the automation
simulations for fixed system parameters and varying system size to those 
for a fixed system size and varying parameters. This scaling is not valid
as is evident from the results presented above. This signals a breakdown
of the reaction-diffusion equation model for the fast reaction rates
considered in this subsection. Here several reactions were carried out for
each diffusion step giving rise to strong local fluctuations due to
reactions which could not be smoothed by diffusion. 

If one considers dynamics in the periodic regime ($\kappa_2=12.8$
described above) and decreases the diffusion coefficient or increases 
the system size, destruction of phase coherence is observed. Such 
destruction of phase coherence on large scales is well known in
oscillatory media and gives rise to phase turbulence.\cite{kur} The 
present results for the chaotic regime show that even though the 
dynamics in smalll correlated regions is locally chaotic, phase 
coherence on the chaotic attractor is not maintained on long 
scales, a result that parallels that in oscillatory media. Thus, 
our stochastic reactive lattice-gas simulations of the autocatalator 
do not show a robust global periodic attractor that survives as the 
system size increases. In certain parameter regimes, such robust 
global attractors were seen in deterministic systems of coupled R\"{o}ssler 
ODEs.\cite{bcm}
 
\section{Space and Time Scales}  \label{scales}
A number of aspects of fluctuations and their effects on chaotic dynamics
which were described above depend on the space and time scales involved in
the automaton simulations. In order to use these results to interpret the
behavior of physical systems one must establish a link between the scales
in the automaton simulations and those that arise in real systems. Any
connection of this type must take into account the characteristics of a
specific system so we shall simply present a few order-of-magnitude
estimates for representative cases to illustrate the points. 

Consider a liquid-state (aqueous) reacting system where the reactants and 
products are small molecules. For a $0.01$M solution, a volume of 
solution with a linear dimension of $0.01 \mu$ contains about three 
reactant molecules, a typical occupancy of an automaton node in our 
simulations. Thus, 
we may associate a linear dimension of $\ell_a=0.01 \mu$ with the automaton 
lattice spacing. Taking the molecular diffusion coefficient to be 
$D \sim 10^{-5}\; {\rm cm}^2/{\rm s}$, the average time for a molecule to
move by diffusion to a neighboring cell is  $\ell_a^2/D=\tau_a \sim 0.1\;
\mu {\rm s}$. This time can be associated with an automaton time step. 

Next, we must consider the time scale of the reactive events in comparison
with that for the non-reactive events, those that are largely responsible 
for the diffusion process, in order to see under what circumstances these 
scales may be applied. For this puropose we may use the average cycle
time observed in the global dynamics. 
In the simulations presented in this paper the average cycle time was of
order $t_c \sim 10^3$ time steps, and this corresponds to 
$t_c \sim  0.1\; {\rm ms}$, a rather fast chemical relaxation time for
most typical oscillatory chemical reactions, but not unrealistic for
reactions under certain conditions. 

More usually the cycle time of
many laboratory examples of oscillatory reactions lie in the range 
$1-100\; {\rm s}$. One may then work backward and associate a time of 
$10^{-3}\; {\rm s}$ with an automaton time step for a cycle time of
$t_c=1\; {\rm s}$ from which it follows that the automaton cell dimension
is roughly $10^{-4}\; {\rm cm}$ if the same value of the diffusion
coefficient is used. In a cell volume of this size there are about $10^7$
reactant 
molecules so that the fluctuations in the automaton are about $10^3$ times
larger than in a system of this type. Of course, these estimates serve to
confirm that in usual circumtances, on the space and time scales of 
many macroscopic experiments, internal fluctuations play only a minor
role. However, they are crucial in homogeneous nucleation processes and
near bifurcation points they can play an essential role.

More importantly, the above calculations and estimates show that under not
very extreme conditions, even less extreme if diffusion coefficients are
smaller than our typical value, one may observe the consequences of the
interactions between internal fluctuations and periodic or chaotic 
dynamics in the spatial domain.

\section{Conclusion} \label{conc}
The reactive lattice-gas automaton provides a description of the 
system at a mesoscopic level. On the smallest scales one has 
individual reactive 
events and random-walk dynamics. Of course, in general, at this level one must 
regard the particles as fictitious, representing some collection of real 
particles. However, one may consider physical circumstances where the particle 
numbers in the automaton correspond closely to those of a real system. In 
this case, the dynamics may be considered to be a model of real collision 
dynamics. At the coarse-grained level one observes a noisy version 
of the periodic or chaotic attractor.  At this mesoscopic 
level one has interplay between the local periodic or chaotic dynamics, 
internal fluctuations and diffusion to give rise 
to the observed spatiotemporal structure.

In the chaotic regime internal fluctuations are amplified by the dynamcis and 
give rise to amplitude fluctuations in the coarse-grained concentration 
variables. Provided the diffusion length is large enough the system 
maintains phase coherence over all space and the phase space trajectory of 
the globallly-averaged concentrations lies on a noisy periodic attractor. 
In the periodic regime fluctuations from the deterministic limit cycle 
exhibit linear growth and a noisy limit cycle is also observed in 
this periodic regime. This behavior is consistent with earlier 
considerations of the noisy dynamcis of coupled map lattices.\cite{bohr} 
If the diffusion length is sufficiently small then the chaotic 
system will no longer be able to maintain phase 
coherence and phase turbulence will ensue, similar to that seen in oscillatory 
media.

The reactive lattice-gas automaton is a synchronously-updated,
probabilistic cellular automaton and thus shares some of the features that
were seen to be essential in the observation of non-trivial global
attractors.\cite{chate,bcm} However, the well-stirred dynamics for large
systems was seen to closely approximate that of the continuous-time mass
action ODE. One may show in this limit that the mean field automaton
equations are simply an Euler discretization of the mass action rate law.
The time step in this discretization is controlled by the scale factor $h$
introduced in (\ref{cmax}) and its small size insured that the automaton
dynamics in this limit yielded results comparable to the solution of the
underlying ODE. The reative lattice-gas model is intrinsically
statistical in nature and encorporates the effects of internal
fluctuations. Thus, there are a number of subtle issues involving the
passage between disctete-time and continuous-time dynamics, and 
their implications for systems with broken time translation symmetry, 
that merit further investigation in the context of these models. 

\vspace{0.2in} \noindent
{\bf Acknowledgements}: This work was supported in part by a grant from
the Natural Sciences and Engineering Research Council of Canada and a by 
a Killam Research Fellowship (R.K.).

\begin{figure}[htbp]
\begin{center}
\leavevmode
\end{center}
\caption{The six panels of this figure show the local concentrations of
species $U$ on a $512 \times 512$ lattice for various times within 
one cycle on the global attractor. Time 
increases from left to right and top to bottom. The concentration is 
coded by gray shades with black high concentration and white 
low concentration of $U$.}
\label{st512}
\end{figure}

\end{document}